\documentclass[final,5p,times,twocolumn,letterpaper]{elsarticle}

\usepackage{natbib}
\usepackage{cancel}
\usepackage{pifont} 
\usepackage{amsthm}         
\usepackage{amsmath} 	   
\usepackage{amssymb}       
\usepackage{amsfonts}
\usepackage{graphicx} 	    
\usepackage{verbatim}
\usepackage{epsfig,bbm}
\usepackage{color}
\usepackage{colortbl}
\usepackage{float}
\usepackage{multirow}

\definecolor{babyblue}{rgb}{0.54, 0.81, 0.94}
\definecolor{corn}{rgb}{0.98, 0.93, 0.36}

\begin{document}

\begin{frontmatter}

\title{Entropy, Black Holes, and the New Cyclic Universe}


\author[add1,add2]{Anna Ijjas }
\author[add0]{Paul J. Steinhardt\corref{cor1}}
\ead{steinh@princeton.edu}
\cortext[cor1]{Corresponding author}

\address[add1]{Max Planck Institute for Gravitational Physics (Albert Einstein Institute), 30167 Hanover, Germany}
\address[add2]{Gottfried Wilhelm Leibniz Universit{\"a}t, 30167 Hanover, Germany}
\address[add0]{Department of Physics, Princeton University, Princeton, NJ 08544, USA}

\date{\today}

\begin{abstract}
We track the evolution of entropy and black holes in a cyclic universe that undergoes repeated intervals of expansion followed by slow contraction and a smooth (non-singular) bounce.  In this kind of cyclic scenario, there is no big crunch and no chaotic mixmaster behavior.  We explain why the entropy following each bounce is naturally partitioned into near-maximal entropy in the matter-radiation sector and near-minimal in the gravitational sector, satisfying the Weyl curvature conditions conjectured to be essential for a  cosmology consistent with observations.   As a result, this kind of cyclic universe can undergo an unbounded number of cycles in the past and/or the future.    
 \end{abstract}



\end{frontmatter}

\noindent
{\bf Introduction.}   Much of the  research on the early universe focuses on trying to explain the homogeneity, isotropy and spatial flatness of the universe, as observed in the cosmic microwave background (CMB) maps of the last scattering surface. However, an equally important but perhaps more vexing issue is explaining the entropy distribution.   

The problem is that, emerging from a big bang in which gravity is strongly coupled and quantum fluctuations of stress-energy and spacetime  are both large, the natural expectation is that  the total entropy should be nearly maximal and equally distributed among both stress-energy and gravitational degrees of freedom.   Consequently, the Weyl curvature $C_{\lambda \mu \nu \xi}$, the traceless part of the Riemann curvature tensor that includes the entropy contributions of tidal fields and gravitational waves,  should be nearly divergent.    However, the  observed entropy distribution on the last scattering surface is puzzlingly different, as Penrose has emphasized \cite{Penrose:1979azm,Penrose:1988mg}.   The total entropy is exponentially smaller than the maximal possible and strangely partitioned:  nearly maximal  in the hot matter-radiation sector (since this sector is near thermal equilibrium) and totally negligible in the gravitational sector.  Furthermore, since $C_{\lambda \mu \nu \xi}$ vanishes for a homogeneous and isotropic background, the Weyl curvature is negligibly small instead of being nearly divergent.

The entropy puzzle is logically distinct from explaining the homogeneity and isotropy
 since it is possible in principle to have a universe that satisfies the latter conditions but with a completely different partition of entropy.    In fact, inflation, which was introduced to smooth and flatten the universe, makes the entropy problem exponentially worse.   In order for inflation to start, the universe must go from the big bang to a state dominated by a nearly uniform field (with all other matter-radiation and gravitational components being negligible).  This requires  an exponentially smaller and, therefore, more fine tuned initial  entropy than is required to explain the last scattering surface without inflation \cite{Penrose:1988mg}.  Similarly, a bouncing cosmology with a singular bounce, in which the transition from contraction to expansion occurs near the Planck density, does not resolve the entropy problem either, since quantum gravity effects near the bounce would cause the Weyl curvature to diverge in that case, too.   

In this paper, we explore a third possibility: cyclic bouncing cosmologies that entail a period of slow contraction \cite{Cook:2020oaj} followed by a gentle (non-singular) bounce to an expanding phase  that occurs well below the Planck density, as has been proven to be possible in \cite{Ijjas:2016vtq,Ijjas:2017pei}.  It has recently been demonstrated  \cite{Ijjas:2021gkf} that slow contraction (see also \cite{Khoury:2001wf,Erickson:2003zm}) is a powerful mechanism for smoothing and flattening the universe even if the conditions at the start of the contraction phase are far from Friedmann-Robertson-Walker (FRW).   Here we focus on the entropy issue, tracking the entropy evolution through various stages of each cycle.  This includes constructing a conformal diagram to take account of the fact that, between the last scattering surface and today, most of the entropy is converted into gravitational entropy in the form of black holes.   The central result is that there is no entropic limit on how many cycles there could have been in our past or will be in our future.

\noindent
{\bf Slow contraction and non-singular bounces}
Bouncing models based on slow contraction and non-singular bounces require familiar elements: ordinary matter, dark matter,  radiation, and a  scalar field $\phi$ with potential $V(\phi)$ that is coupled to itself and gravity and that evolves throughout according to classical equations of motion in the standard $(3+1)$ dimensions.  (We note that the bouncing models discussed here do not require extra-dimensions, branes or any phase in which quantum gravity dominates.)  

Slow contraction refers to a phase in which the Friedmann-Robertson-Walker (FRW)  scale factor $a(t)$ scales as 
$(\Delta t)^{1/\varepsilon}$ where $\varepsilon > 3$ and $\Delta t$ is the time remaining before a bounce.
The phase can be generated by something as simple as 
a canonical scalar field evolving down an exponentially steep negative sector of its potential where $V(\phi)
\sim {\rm exp} (\phi/m)$.   There exists then an attractor solution in which the equation of state of the scalar field, 
$\varepsilon \equiv 3/2 (1+p/\rho) \gg 3$, where $p$ is the pressure and $\rho$ is the energy density, converges to $1/2 m^2$.  The attractor solution corresponds  a value of $\varepsilon$ that can be quite large; for example, for $m=0.1$ or 0.01, $\varepsilon = 50$ or 5000.  
Consequently, 
the scalar field energy density ($\propto 1/a^{2 \varepsilon}$)  grows rapidly to dominate over all other forms of stress-energy, the spatial curvature or the anisotropy).   The rapid growth is due to gravity:  the scalar field kinetic energy is gravitationally blue-shifted at an extraordinary rate when the universe is contracting.

In this way, slow contraction resolves the homogeneity, isotropy, and flatness problems.  Extensive numerical relativity studies \cite{Ijjas:2020dws,Ijjas:2021gkf,Ijjas:2021wml} have shown that slow contraction is a remarkably powerful, rapid and robust smoother and flattener that is surprisingly insensitive to initial conditions and avoids quantum runaway effects that  lead to the multiverse problem.  For a wide range of cases, the smoothing and flattening is complete by the time $a(t)$ shrinks by a factor ${\cal O}(10)$ or less.  

Consequently, by the time the bounce occurs, spacetime is nearly FRW, the Weyl curvature is negligible, and any entropic contributions to the total stress-energy are negligible compared to the homogeneous scalar field density.  Assuming a bounce mechanism that  is non-singular and occurs at energy densities well below the Planck density (as in the examples in Ref.~\cite{Ijjas:2016vtq}), these homogeneous and isotropic conditions should remain  after the bounce.   At that point, contraction is replaced by expansion; the Hubble parameter changes sign; and, as a result, the gravitational blue-shifting of the scalar field energy density comes to an end. Now, through the interactions of the scalar field with other fields, the same sort of reheating mechanism envisaged for an inflaton scalar field can lead to the rapid conversion of the now-dominant scalar field energy density to a thermal distribution of matter and radiation.  Since the decay occurs uniformly over space, homogeneity and isotropy are maintained.  
  
The net result is  a spatially flat  FRW universe with negligible Weyl curvature;  total entropy much less than maximal; and, the entropy that does exist is partitioned into nearly maximal entropy in the thermally equilibrated matter-radiation sector and negligible entropy in the  gravitational and scalar field sectors.  

This single bounce scenario is promising, but it does not include any explanation for what preceded the contraction phase or what the ultimate fate of the universe will be.   A cyclic universe is appealing because it  potentially provides a more complete and more predictive model.  Entropy has famously been a problem in earlier types of cyclic models; but here we show how slow contraction and a non-singular bounce lay those problems to rest.

\noindent
{\bf Cyclic models with slow contraction and a non-singular bounce.}
In cyclic models of this type, the scalar field plays multiple roles as it evolves back and forth along its potential $V(\phi)$. It is at one stage the source of the dark energy that drives the current accelerated expansion (and its equivalent in other cycles); at another stage, it is the component responsible for the transition from accelerated expansion to slow contraction;  at a third stage, it is the driver of the transition from contraction to expansion (the bounce);  and, through the decay of scalar field stored energy, the origin of the hot matter-radiation that dominates the universe after the bounce and for the first 9 Gy following (before the scalar field acting as dark energy takes over).  Microphysical models exemplifying these stages can be found in earlier papers (see \cite{Ijjas:2019pyf} and references therein. For the purposes here, it suffices to know that the sequence of phases is possible.

\begin{figure}[t]
\includegraphics[width=3.00in,angle=-0]{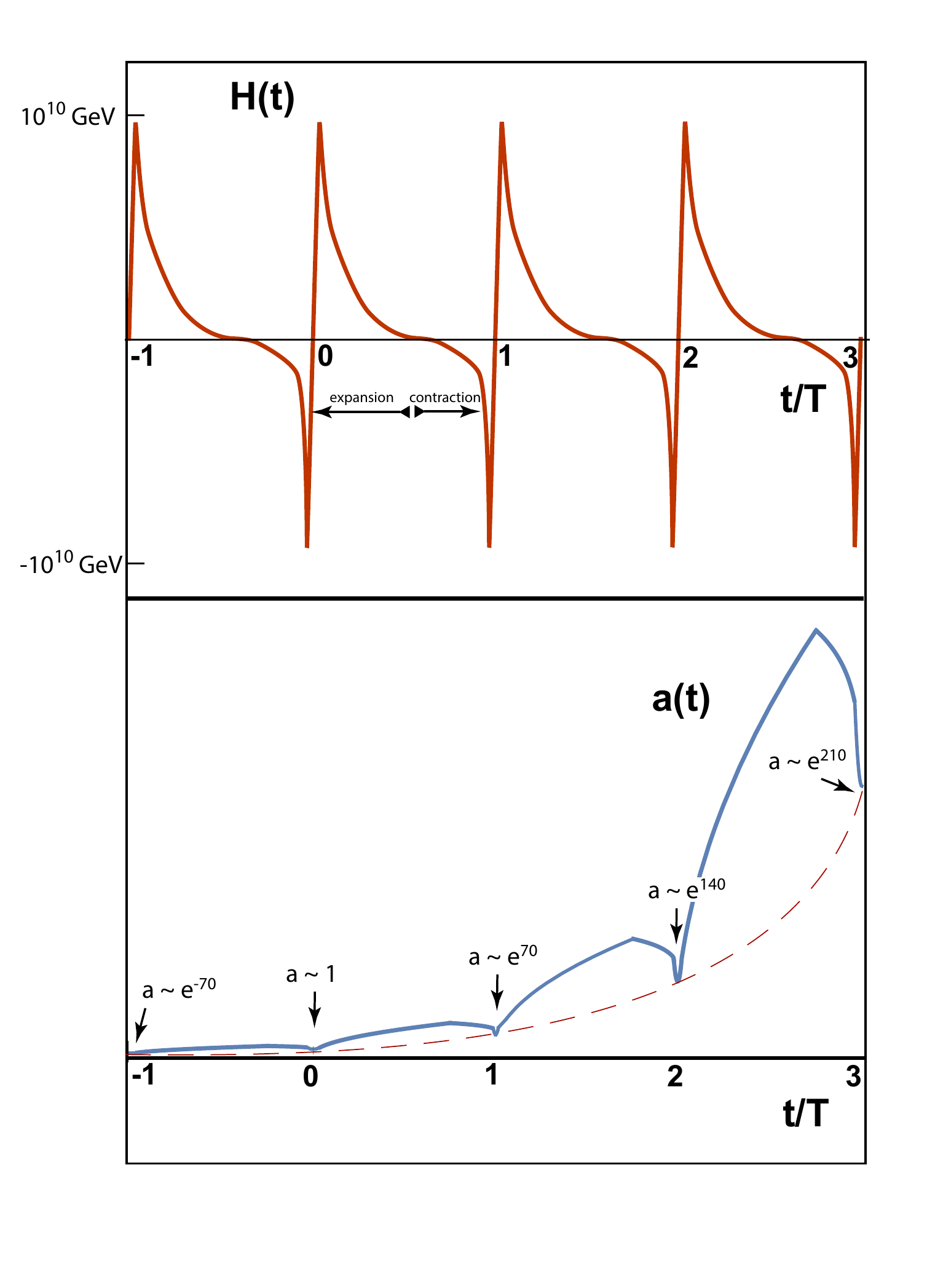}
\caption{ \label{Figure0} In the cyclic model based on slow contraction and a non-singular bounce, the Hubble parameter $H(t)$ oscillates (upper panel) between positive values (during expansion) and negative values (during contraction) with an amplitude of $\sim 10^{10}$~GeV over a period $T$.  The scale factor $a(t)$, by contrast, goes through long periods of expansion followed by shorter periods of contraction (lower panel).  The net result is an exponential increase in $a(t)$  over the course of each cycle, producing an on-average de Sitter-like expansion (dashed curve in lower panel) over many cycles. This figure is adapted from Ref.~\cite{Ijjas:2019pyf}.}
\end{figure}

The striking and distinctive feature compared to other types of cyclic models is that the evolution of the Hubble parameter $H(t) \equiv d \ln a/dt$ is periodic, but the evolution of $a(t)$ is not \cite{Ijjas:2019pyf,Biswas:2008kj,Biswas:2010si}.  Instead, $a(t)$ increases by a substantial exponential factor from cycle to cycle.   See Fig.~\ref{Figure0}.

To be more precise: Over the course of a cycle, $a(t)$ increases by 60 e-folds during the radiation- and matter-dominated epochs (assuming a reheat temperature of $\sim 10^{15}$~GeV after the bounce), and by an additional   $N_{\rm DE}$ e-folds during the dark energy dominated epoch, where $N_{\rm DE}H_0^{-1}$ is the duration of  dark energy epoch and 
$H_0^{-1}$ is the present value of the Hubble parameter.    During the subsequent slow contraction and bounce phases, the decrease in $a(t)$ is ${\cal O}(1)$; that is, negligible by comparison to the $(60+N_{\rm DE})$ e-folds of increase during the expanding phase.  The extreme asymmetry between the large amount of expansion versus the tiny amount of contraction of $a(t)$ is due to the fact that $\varepsilon$ is $O(1)$ or less during the expansion phase but $\gg 1$ during the contraction phase.  Over the course of many cycles, the result is an on-average de Sitter-like expansion over many cycles with an effective de Sitter Hubble parameter of $H_{\rm deS} = (N_{\rm DE}+60) H_{0}^{-1}/(N_{\rm DE}+1)$ that is set by the energy density and dynamics of the dark energy density.  

The scale factor as shown in Fig.~\ref{Figure0} is approximately of the form
\begin{equation}
\label{scale}
a(t) = P(t) \;  {\rm exp} \left(\Big(N_{\rm DE}+60\Big) \frac{t}{T} \right),
\end{equation}
where $P(t+ T) = P(t)$  is a periodic function with period $T\approx ( N_{DE}+1)H_0^{-1}$ and the exponential factor accounts for the exponential increase in $a(t)$ from cycle to cycle.  Without loss of generality, we can  choose $t=0$ to correspond to the moment during the most recent bounce when $H$ passed through zero (so 
$\dot{a}(0)=0$) and  normalize the scale factor so that $a(0)=1$.  In this case,  $P(0)=P(T)=1$ and $\dot{P}(0)=\dot{P}(T)=-(N_{\rm DE}+60)/T$.  The precise form of
$P(t)$ can be computed from knowing how $a(t)$ varies with time during each phase.   Note that $a(t)$ is, by definition, a conformal factor.  Consequently, even though its value is exponentially smaller in earlier cycles compared to today, this has no effect on physical observables, like the temperature, density, expansion rate, etc.

The  corresponding   Hubble parameter varies periodically:
\begin{equation} 
\label{Hubble}
H(t) = (d \; {\rm ln} P(t)/dt) + (N_{\rm DE}+60) /T.
\end{equation}
As shown in Fig.~\ref{Figure0}, $H(t)$ oscillates between positive and negative values, passing through zero twice each cycle.  The time-independent second term in Eq.~(\ref{Hubble}) is small and positive; the large range in $H(t)$ is due to the first term, which runs from $\gg (N_{\rm DE}+60) /T$ to $ \ll -(N_{\rm DE}+60) /T$ over the course of a cycle.

As for the Hubble radius, $|H^{-1}|$,  it grows after each bounce
from a microscopic size (say, $ \sim 10^{-25}$~cm) to the  $10^{28}$~cm Hubble radius observed today.  It remains roughly constant during the dark energy dominated (accelerated expansion) phase. Then the Hubble radius shrinks
 back to  microscopic size ($\sim 10^{-25}$~cm) during slow contraction and the process repeats after the bounce.
The enormous decrease in the Hubble radius during slow contraction is possible even though a(t) shrinks very little because the Hubble radius is proportional to $a(t)^{\varepsilon}$ according to the Friedmann equation, and $\varepsilon$ can be quite large ({\it e.g.} $\ge 50$).  

\noindent
{\bf Four roadblocks to overcome.}
Cyclic models in which $a(t)$ varies periodically and/or approaches zero at the transition from contraction to expansion  encounter one or more of the following four roadblocks, each of which is avoided in the cyclic model considered here: 

\noindent
 {\it Cosmic singularity}, which leads to the Weyl curvature problem:  avoided because the energy density is at all times well below the Planck density and $a(t)$ does not approach zero.
 
 \noindent
 {\it Chaotic mixmaster oscillations} triggered by anisotropy, which  destroy homogeneity and isotropy: avoided because a scalar field with $\varepsilon >3$ dominates the anisotropy during the contraction phase, altogether blocking mixmaster oscillations.

 \noindent
{\it Big crunch}, which causes black holes formed in the preceding cycle to merge and disrupt the bounce:  avoided because  $a(t)$ and, hence, the distance between black holes,  does not decrease significantly during the slow contraction phase.  And last but not least:

\noindent
{\it Entropy density build-up} from cycle to cycle that limits the number of cycles (a.k.a.~the Tolman entropy problem, see Ref.~ \cite{1932PhRv...39..835T}):  this is the subject of the remainder of this paper.

 \vspace{0.1in}
\noindent
{\bf  The cyclic conformal diagram and the evolution of black holes.} 

Entropy evolution from cycle to cycle can be envisaged by constructing the conformal diagram for this scenario.  As is standard with conformal diagrams, we first consider the sequence of stages using  
({\it barred})  conformal time and space coordinates $(\bar{\eta}, \, \bar{\chi})$, and then rescaling  to new ({\it unbarred}) coordinates $(\eta,\; \chi)$ chosen such  that  they span a finite range and preserve the condition that null lines are oriented at $\pm 45$~degrees.

In the cyclic model considered here, there is no crunch -- in fact, $a(t)$ increases from bounce to bounce --  and quantum gravity effects are negligible throughout.  Therefore, $a(t)$ is always positive and well-described by classical equations of motion.    As a result, the conformal time can be straightforwardly computed from the relation 
\begin{equation} \label{eta}
d \bar{\eta}  \equiv dt/a(t)
\end{equation} 
 using the expression for $a(t)$ with FRW time coordinate in Eq.~(\ref{scale}), even 
if the classical equations deviate from Einstein gravity near the bounce.  

\begin{figure}[t]
\center{\includegraphics[width=3.25in,angle=-0]{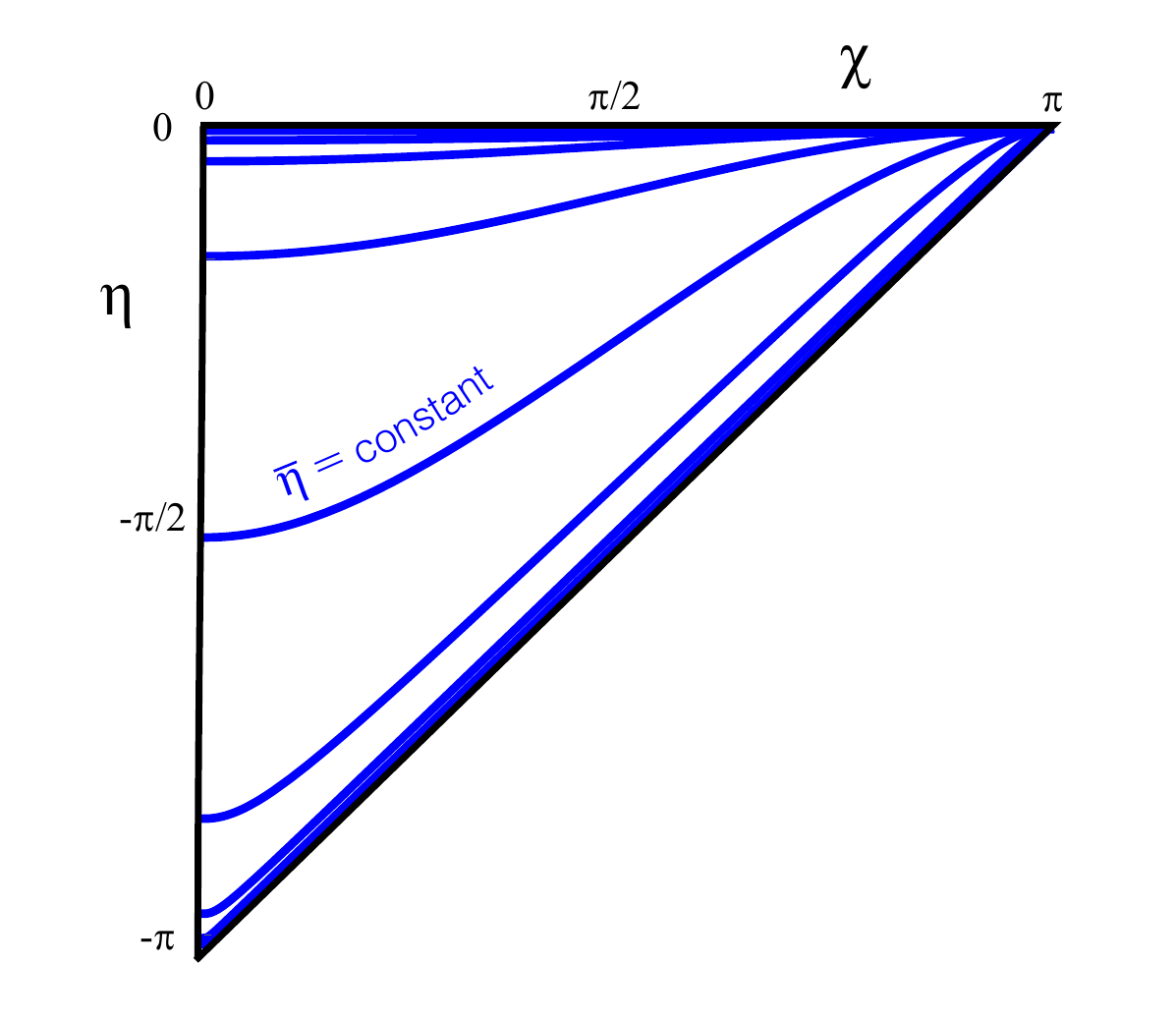}} 
\caption{The conformal diagram for a cyclic cosmology based on slow contraction and non-singular bounces.    The thick blue curves (centered on curves of constant $\bar{\eta}$) correspond to the 
the slow contraction and bounce phases during which $a(t)$ (and, hence, $\eta$)  changes negligibly.  The overall conformal diagram bears a superficial resemblance to that for flat de Sitter space, but here $H(t)$ varies over a wide span of values that range from positive to negative. }
 \label{Figure1}
\end{figure}

Because of the exponential factor in Eq.~(\ref{scale}) that describes the envelope for the evolution of $a(t)$, the result for  $\bar{\eta}$ bears a close resemblance to the result for flat de Sitter space with $H_{\rm deS}=(N_{\rm DE}+60)/T$.   Namely, $\bar{\eta}$ spans a semi-infinite range that can be chosen to be $0 > \bar{\eta} > -\infty$; similarly, $+ \infty> \chi > 0$.  Consequently, one can use the same transformation as in the true flat de Sitter case  to convert from the ({\it barred}) 
to  the conformally equivalent ({\it unbarred}) coordinates $(\eta, \; \chi)$ related by
\begin{equation} \label{compact}
\bar{\eta} = \frac{ \sin \eta}{\cos \eta + cos \chi} \; {\rm and} \; \bar{\chi} = \frac{sin \chi}{\cos \eta + cos \chi} 
\end{equation}
that span a finite range $0> \eta > -\pi$ and $\pi > \chi > 0$.  
Then the conformal diagram takes the form of Fig.~\ref{Figure1}.  Space-like infinity corresponds to the upper right corner of the triangle; past time-like infinity to the lower left hand corner; and past null-like infinity to the diagonal.

Although the overall causal structure is formally similar to flat de Sitter, the cosmic scenario is fundamentally different. For example, the Hubble parameter $H(t)$ is positive and time-independent in flat de Sitter, but $H(t)$ traverses a wide range from highly positive to highly negative during each period of the cyclic model.   Also, flat de Sitter has constant positive accelerating expansion, $\ddot{a}/a>0$, whereas the cyclic model may not include any interval with this property.  Hence, it is important to include demarcations in the conformal diagram of the cyclic model that indicate these and other differences.

Fig.~\ref{Figure1} includes (blue) curves that indicate the slow contraction and bounce phases.  From one bounce to the next, $a(t)$ increases by an exponential factor, ${\rm exp} (N_{\rm DE}+60)$; therefore, the conformal time interval between bounces, $\Delta \bar{\eta}$, decreases by this same factor from bounce to bounce, according to Eq.~(\ref{eta}).  
That is, in the {\it barred} coordinates ($\bar{\eta}, \; \bar{\chi}$ (which have semi-infinite ranges), the bounces would be represented as semi-infinite horizontal lines of constant $\bar{\eta}$ whose spacings decrease as $\bar{\eta} \rightarrow 0$.  
However, this pattern gets distorted when we re-plot using 
the transformations in Eq.~(\ref{compact}) that convert $(\bar{\eta}, \; \bar{\chi})$ to the finite-range ({\it unbarred})  coordinates $(\eta,\; \chi)$ shown in 
Fig.~\ref{Figure1}.   The semi-infinite horizontal lines of constant $\bar{\eta}$  are transformed to finite curves that meet at space-like infinity and compress the intervals between bounces approaching past null-like infinity (the diagonal).  An artifact of the transformation in Eq.~(\ref{compact}) is that intervals between bounces in the middle range ($\eta \sim -\pi/2$) appear to be the widest even though they actually decrease monotonically going from bottom to top in the physical ({\it barred}) coordinate $\bar{\eta}$.  

Without loss of generality, we can choose the blue curve that intersects the vertical axis near $\eta=- \pi/2$ to represent the most recent bounce that occurred 13.8 Gya.  The blue curves below and above then represent the bounces in our past and future, respectively.  Only a few bounces are shown on either side to avoid cluttering the diagram.

The scale factor $a(t)$ (and, hence, $\bar{eta}$) changes negligibly during the slow contraction and bounce phases combined compared to the interval between one bounce and the next (${\rm exp} (N_{\rm DE}+60)$).   This comparatively small change in $\bar{\eta}$ is represented symbolically by the thicknesses of the blue curves.  If the thicknesses were drawn to proper quantitative scale,  the curves would be exponentially thinner.  For pedagogical purposes, we have made them thick to remind the reader that they represent not just a single instant of time, but the combination of the slow contraction and non-singular bounce stages.

(N.B. A very similar looking conformal diagram for a different kind of cyclic model appears in \cite{Steinhardt:2004gk}; in that case, the curves correspond to big crunches where  $a(t) \rightarrow 0$, densities and the Weyl curvature diverge, pre-existing black holes merge, and the Hubble radius shrinks to zero; consequently, the conclusions reached in this paper do not apply to that case.) 

Immediately following each bounce, the universe enters a radiation-dominated phase due to the decay of the scalar field energy density.  Spacetime is homogeneous, isotropic and spatially flat due to the preceding slow contraction phase.
At these moments, the matter-radiation is in thermal equilibrium at nearly maximal entropy.   On the other hand,  the gravitational sector entropy and the Weyl curvature are negligible.  The subsequent  expansion of the universe between the bounce and the last scattering surface is nearly adiabatic  and homogeneous, so the distribution of entropy on the last scattering surface is the same.  (Here and in the remainder of this discussion, we consider the evolution of entropy within a comoving volume whose radius is $\sim H_0^{-1}$ using time-slicing on surfaces of constant (FRW) cosmic time.)
  
\begin{figure}[t]
\center{\includegraphics[width=3.0in,angle=-0]{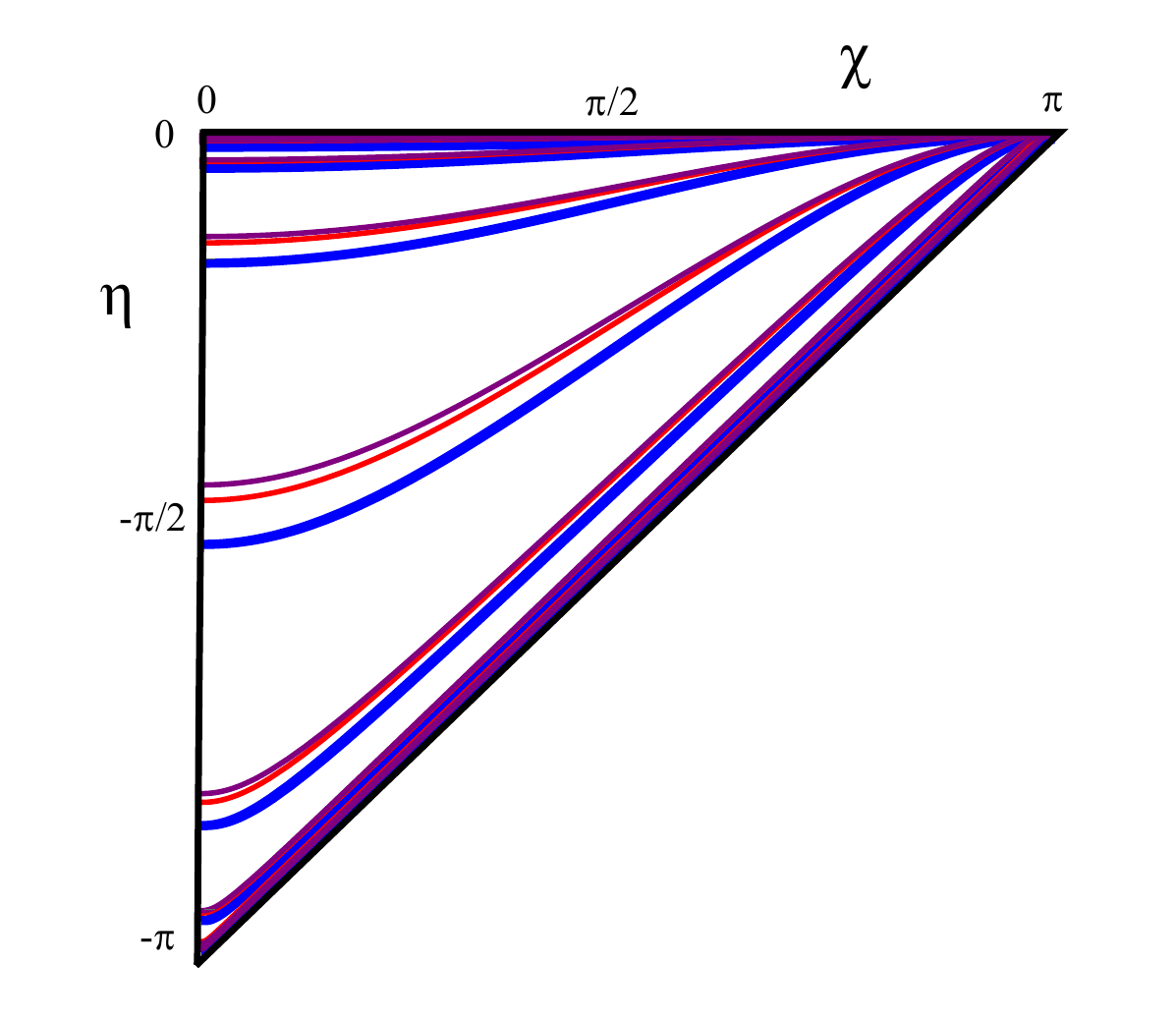}} 
\caption{An enhanced version of the conformal diagram in Fig.~\ref{Figure1} where each blue curve represents the constant $\bar{\eta}$ surfaces corresponding corresponding to the slow contraction and bounce phases.  The interval between each blue curve and the red curve directly above it corresponds to a radiation dominated phase.   Each red curve represents  radiation-matter equality and, interval  between it and the violet curve just above it corresponds to a 
matter dominated phase; each violet curve represents the crossover from matter to dark energy domination (accelerated expansion); which terminates at the next blue curve above it, when the next contraction and bounce phases occur. }
 \label{Figure2}
\end{figure}

In Fig.~\ref{Figure2}, we have added two curves to each interval between bounces that represent the transition from radiation to matter domination (red) and the transition from matter domination to dark energy domination.  As we did for the thickness of the blue curves above, we have for pedagogical purposes exaggerated the spacings between these different curves, which span the first 10 Gy since the last bounce.  If drawn to proper scale, the bounce, matter-radiation equality and dark energy-matter equality curves would be too close to discriminate from one another
 (because the spacing depends the factor by which $a(t)$ (and, hence, $\bar{\eta}$) changes in going from one stage to the next, and that factor is exponentially small compared to the factor by which $a(t)$ changes in from one bounce to the next).

Although the  entropy on the last scattering surface is almost entirely in the matter-radiation sector, an important change occurs as the universe continues to expand and cool.    Non-linear structure forms through gravitational instability and  a combination of irreversible processes that cause the total entropy to increase significantly.  Most of that increase is in the form of Bekenstein-Hawking \cite{Bekenstein:1974ax,Gibbons:1977mu}  entropy associated with supermassive black holes (SMBHs) that have formed since last scattering \cite{Egan_2010}.  Based on current measurements of cosmic parameters and the mass function for SMBHs, a census of the entropy contributions  within the observable universe today  yields $S_{\rm gas \& stars}\sim 10^{81}$k, $S_{\rm photons}\approx S{\rm neutrinos} \approx S_{\rm dark \, matter} \sim 10^{90}k$, 
$S_{\rm baryons} \sim 10^{90}k$, and $S_{\rm SMBH} \sim 10^{105}k$ where $k$ is Boltzmann's constant.  
The black hole contribution overtakes the matter-radiation entropy around $z=10$ \cite{Egan_2010,LPageAndCJagoe}, 
   and, by the present epoch (or its equivalent in other cycles) exceeds the other sources of entropy by a factor $\sim (10^{15}$).

During the remaining dark energy dominated phase, the  entropy density in the form of radiation and black holes are both diminished, but the radiation density decreases faster than the black hole mass density.  Hence, to determine how much entropy from a previous cycle lies within the Hubble radius a cycle later, it suffices to track what happens to the supermassive black holes.
For this purpose, it is reasonable to treat the black holes as moving along comoving worldlines (constant $\bar{\chi}$). 

During the slow contraction phase, the  black hole {\it density} does not change significantly because $a(t)$ only decreases by a factor of a few.  

In cyclic models in which $a(t)$ shrinks significantly during contraction (by more than $10^{30}$ or more, say), spacetime would be filled with supermassive black holes that crunch together, and, in cases where $a(t) \rightarrow 0$ at the bounce, it would be unclear how to treat the passage of black holes through to the expanding phase.   
However, in cyclic models based on slow contraction and non-singular bounces, though, $a(t)$ may only contract by a factor of two or so, as noted above.  Spacetime remains far from singular.  Consequently, mergers do not take place and it is reasonable to suppose that supermassive black holes move along comoving worldlines pass through the bounce freely.  

More precisely, during contraction, any Hubble patch far from the black holes  shrinks to a tiny size (radius~$\sim 10^{-24}$~cm) that is much smaller than the radii of supermassive black holes.  These regions undergo the bounce and reheating process that have been described.   Regions within  the gravitational near-field of the black holes will undergo a different evolution determined by that local field sourced by the black hole.  Although we cannot precisely describe what that is, it is plausible to imagine the black holes continue to exist after the bounce, but now immersed in a cosmic far-field metric that is expanding rather than slowly contracting.  The evaporation time of these black holes  is much longer than a cycle, assuming no accretion.  In fact, the black holes would accrete significant mass during the phases following a bounce.  We presume the net effect is that the black holes can survive indefinitely from cycle to cycle.    

\begin{figure}[t]
\center{\includegraphics[width=3.5in,angle=-0]{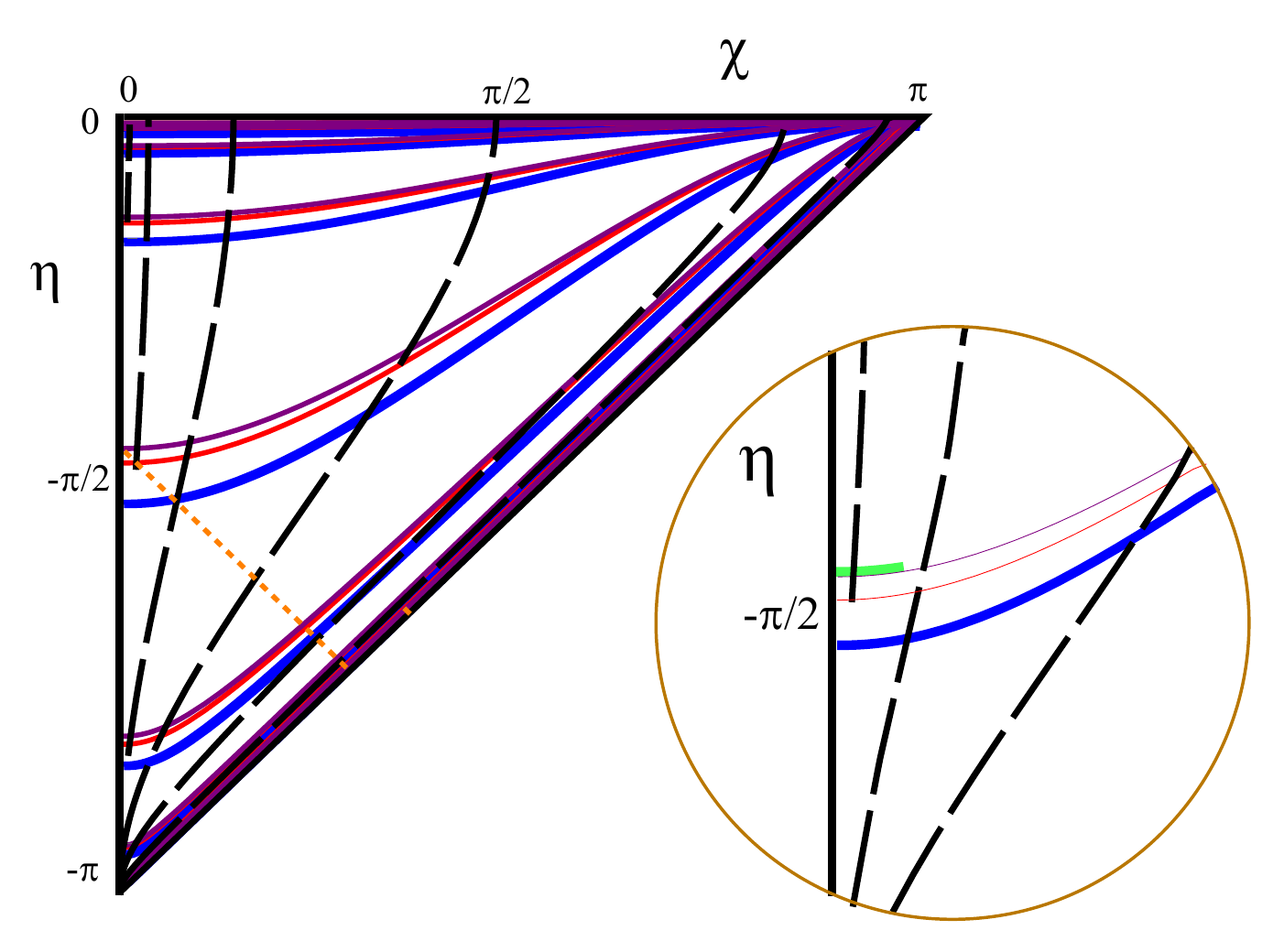}} 
\caption{On the left, the conformal diagram in Fig.~\ref{Figure2} where black dashed curves have been added that represent the worldlines for a sequence of black holes, each formed during a matter dominated phase (one black hole per cycle), each at the same time interval following the most recent bounce, and  each at the same distance from a comoving observer at $\chi=0$.   Also shown in orange is the particle horizon for an observer today ($\eta=\eta_0$) located at $\chi=0$, where  the blue curve that intersects $\chi=0$ near $\eta=\pi/2$ represents the most recent bounce that occurred 13.8 Gya.  Note that black holes formed in earlier cycles are outside the particle horizon of the observer in a current one. In the circle at the right is a blowup showing 
the most recent bounce (blue), the matter-radiation equality (thin red) and dark energy-matter equality (thin purple),and the 
current extent of the observable universe (green).
   Note that black holes formed in earlier cycles lie outside the observable universe today and so do not contribute the entropy census described in the text.}
 \label{Figure3}
\end{figure}

Consequently, the conformal diagram in Fig.~\ref{Figure3} shows the world lines of a representative sequence of black holes continuing along a comoving worldline from cycle to cycle.  In this representative sequence, each black hole formed during a matter dominated phase (one black hole per cycle), each  at the same time interval following the most recent bounce, and  each at the same distance from a comoving observer at $\chi=0$.  Recall that the scaling is not accurate so the figure should be viewed as qualitative; drawn to proper scale, the distances between the curves corresponding to the bounce, the matter-radiation and dark energy-matter equality curves in a given cycle should be exponentially smaller than shown here.   The figure shows that black holes formed in previous cycles exit the particle horizon of today's observer before the most recent bounce and so do not contribute to today's observable black hole density.  Also, the radiation emitted by them is diluted during the dark energy dominated expansion phase and then overwhelmed by the new radiation produced after the bounce, so there is no measurable effect in the next cycle.

Quantitatively, suppose a black hole was at a distance $L$ from a comoving observer's worldline one cycle ago.  For that black hole to lie within that observer's  Hubble radius today, it is necessary that 
 \begin{equation}
L \times (e^{N_{\rm DE}}) \times O(1) \times (e^{60})< H_0^{-1}
\end{equation}
or $L < 0.004$~cm, where the factors correspond to the dark energy expansion, slow contraction, and radiation plus matter expansion epochs between a cycle ago and today.   Equivalently, the volume that evolved into today's Hubble volume of $(10^{28} \; {\rm cm})^3$ occupied only $(0.004 \; {\rm cm})^3$ a cycle ago, so only the entropy and black holes that were within that small volume a cycle ago would be within our Hubble volume today. 

Considerations of entropy naturally lead to considerations of the second law of thermodynamics.  In this case, while we do not claim to explain the arrow of time, there is no apparent violation  of the second law since dissipation is included, heat always flow from hot to cold, and the total entropy of the universe is always increasing. (It is only the entropy within a Hubble radius that exponentially decreases during contraction, and only  because the Hubble radius is shrinking during the contraction phase so that entropy that was within the Hubble radius exits during the slow contraction phase.  This violates no fundamental laws.)  Furthermore, in the limit of an infinite universe, there are an unbounded reservoir of energy and an infinite volume to accommodate the  increasing entropy and increasing number of black holes.  Hence, there is no obvious entropic roadblock.

Nevertheless, it has been suggested recently that an infinitely bouncing universe of the type here violates a conjectured cosmological generalization of the second law for de Sitter-like spacetimes (inspired by a conjectured `central dogma' about black holes) \cite{Susskind:2021yvs}.  The argument, though, rests on holographic reasoning that implicitly assumes null convergence ({e.g., Einstein gravity and the null energy condition), which effectively rules out bounces by fiat, rather than by some truly independent evidence.

 \vspace{0.1in}
\noindent
{\bf  Discussion.}  In sum, each cycle brings the universe to the same conditions after the bounce,  conditions in which:  
\begin{itemize}
\item the total entropy with the Hubble radius  is much less than maximal; 
\item the entropy with the Hubble radius  is partitioned so that it is nearly entirely in the matter-radiation and nearly minimal in the gravitational sector; 
\item the Weyl curvature is negligible after the bounce; and, 
\item matter, radiation, entropy, and black holes from earlier cycles virtually all lie outside the Hubble radius.  
\end{itemize}
Hence, there appears to be no way for an observer making local observations to distinguish one cycle from another, and there appears to be no limit to the number of cycles that may have occurred in the past or that will occur in the future.   
Of course, tracing the cycles going back in time, $a(t)$ shrinks by a large exponential factor between bounces.  To have a truly unlimited sequence of bounces going back in time requires that spacetime be infinite.  In that limit, $a(t)$ acts purely as a conformal factor that can be arbitrarily small in the past or arbitrarily large in the future without a physical effect on a local observer.  
We also note that the new cyclic cosmology can be made geodesically complete, {\it e.g.}, if a Higgs-like scalar drives the background evolution and the scale factor increases from cycle to cycle.  As explained in Ref.~\cite{Bars:2013vba}, the wordlines of massive free-falling bodies can then be extended arbitrarily far backward in proper time.  The derivation (see  Eq.(7) and the ensuing discussion) does not depend on the energy conditions or the precise bounce mechanism.  
 Globally, the total entropy and number of black holes grows without bound but in proportion to the growing volume so as to keep their densities the same after each bounce.

 \vspace{0.1in}
\noindent
{\it Acknowledgements.} 
We thank  R. Penrose, V. Mukhanov, L. Page and M. Chitoto for useful discussions.
The work of A.I. is supported by the Lise Meitner Excellence Program of the Max Planck Society and by the Simons Foundation grant number 663083.
P.J.S. is supported in part by the DOE grant number DEFG02-91ER40671 and by the Simons Foundation grant number 654561.

\bibliographystyle{apsrev}
\bibliography{EntropyNewCyclic}

\end{document}